\documentclass[aps,pra,reprint,superscriptaddress]{revtex4-1}

\usepackage{graphicx}
\usepackage{amsmath}
\usepackage{amssymb}

\DeclareMathOperator{\Pf}{Pf}

\begin{document}

\title{Quasiparticles as Detector of Topological Quantum Phase Transitions}

\author{Sourav Manna}
\affiliation{Max-Planck-Institut f\"ur Physik komplexer Systeme, D-01187 Dresden, Germany}

\author{N. S. Srivatsa}
\affiliation{Max-Planck-Institut f\"ur Physik komplexer Systeme, D-01187 Dresden, Germany}

\author{Julia Wildeboer}
\affiliation{Max-Planck-Institut f\"ur Physik komplexer Systeme, D-01187 Dresden, Germany}
\affiliation{Department of Physics, Arizona State University, Tempe, AZ 85287, USA}
\affiliation{Department of Physics and Astronomy, University of Kentucky, 505 Rose Street, Lexington, KY 40506, USA}

\author{Anne E. B. Nielsen}
\altaffiliation{On leave from Department of Physics and Astronomy, Aarhus University, DK-8000 Aarhus C, Denmark.}
\affiliation{Max-Planck-Institut f\"ur Physik komplexer Systeme, D-01187 Dresden, Germany}

\begin{abstract}
A number of tools have been developed to detect topological phase transitions in strongly correlated quantum systems. They apply under different conditions, but do not cover the full range of many-body models. It is hence desirable to further expand the toolbox. Here, we propose to use quasiparticle properties to detect quantum phase transitions. The approach is independent from the choice of boundary conditions, and it does not assume a particular lattice structure. The probe is hence suitable for, e.g., fractals and quasicrystals. The method requires that one can reliably create quasiparticles in the considered systems. In the simplest cases, this can be done by a pinning potential, while it is less straightforward in more complicated systems. We apply the method to several rather different examples, including one that cannot be handled by the commonly used probes, and in all the cases we find that the numerical costs are low. This is so, because a simple property, such as the charge of the anyons, is sufficient to detect the phase transition point. For some of the examples, this allows us to study larger systems and/or further parameter values compared to previous studies.
\end{abstract}

\maketitle

\section{Introduction}

Describing physical systems in terms of phases allows us to focus on key properties rather than the full set of microscopic details. Quantum phase transitions take place at zero temperature, when a control parameter, such as the magnetic field strength, is varied \cite{Sachdev1}. In conventionally ordered phases, quantum phase transitions can be characterized by a local order parameter, arising from the broken symmetry of the system, but this approach breaks down for the case of topologically ordered systems \cite{Wen5}. A further complication arises because strongly correlated quantum many-body systems are demanding to study numerically. Density matrix renormalization group investigations are usually limited to one-dimensional systems or quasi two-dimensional systems such as ladders and thin cylinders \cite{Pollmann1}, and many systems that may harbor topologically ordered phases cannot be studied with large-scale quantum Monte Carlo, due to the sign problem \cite{Julia1}.

Different probes have been developed to detect topological phase transitions, such as ground state degeneracy \cite{WenNiu}, many-body Chern number \cite{Haldane1,Wu1,Hatsugai1}, spectral flow \cite{Mudry1,Sheng1,Regnault}, entanglement spectrum \cite{Regnault,Thomale1,Hermanns1,Sterdyniak1}, topological entanglement entropy \cite{Wen1,Preskill1,Balents1}, and fidelity \cite{zanardi}. The first three assume particular boundary conditions. The entanglement based probes have been tested for regular structures in two dimensions, but it is not clear how and whether they can be applied in highly irregular systems. Fidelity cannot be used if the Hilbert space itself changes as a function of the parameter. There are hence systems that cannot be handled currently. In addition, it is desirable to find probes that are less costly numerically. There is hence a strong demand for identifying further probes.

Here, we show that quasiparticles are an interesting tool to detect topological quantum phase transitions. It is well-known that topologically ordered systems can host anyons, and their properties define the topological phase. Anyons are quasiparticles that are neither fermions nor bosons, and this can be seen from the braiding statistics. They can also have fractional charge. Both anyonic braiding properties and fractional charge have been confirmed in numerical studies \cite{AroSchWil,BonGurNay,Braiding,Nielsen1,macaluso}. Here, we propose to use quasiparticles to detect phase transitions that happen when a parameter in the Hamiltonian is varied. Our starting point is to modify the Hamiltonian locally to generate quasiparticles at well-defined positions in the ground state. In the simplest case, this can be done by adding a potential, while in more complicated systems, it may require some ingenuity \cite{storni}. We then study the properties of the quasiparticles as a function of the parameter. When the two phases do not support the same set of quasiparticles, a change is seen at the phase transition. The method can be applied for all types of anyons, as long as there is an appropriate way to create the anyons, and it does not require a particular choice of boundary conditions or a particular lattice structure. The method therefore also applies to, e.g., disordered systems, fractals, and quasicrystals.

We test the method on concrete examples, namely phase transitions happening in a lattice Moore-Read model on a square lattice and on a fractal lattice, in an interacting Hofstadter model in the presence and in the absence of disorder, and in Kitaev's toric code in a magnetic field. Among these models, we include cases, for which the phase transition point is already known, since this allows us to compare with other methods and check the reliability of the anyon approach. For all these examples, we find that it is sufficient to compute a relatively simple property, such as the charge of the anyons, to determine the phase transition point. The computations can therefore be done at low numerical costs.

For the Moore-Read model on a square lattice, e.g., a large speed up is found compared to previous computations of the topological entanglement entropy, and this enables us to determine the transition point much more accurately. For the interacting Hofstadter model, we only need two exact diagonalizations for each data point, which is much less than what is needed to compute the many-body Chern number. Finally, for the model on the fractal, we do not know of other methods that could be used for detecting the phase transition.

\section{Lattice Moore-Read model}

The Moore-Read state is a trial wavefunction to describe the plateau at filling factor $5/2$ in the fractional quantum Hall effect \cite{Read1}, and it supports non-Abelian Ising anyons \cite{BonGurNay}. In this section, we investigate phase transitions that happen in lattice versions of the Moore-Read state on two different lattices as a function of the lattice filling.

\subsection{Moore-Read model on a square lattice}

We investigate a model with a particular type of lattice Moore-Read ground state, which was shown in \cite{Nielsen2}, based on computations of the topological entanglement entropy $\gamma$, to exhibit a phase transition as a function of the lattice filling with the transition point in the interval $[1/8,1/2]$. A more precise value was not determined because $\gamma$ is expensive to compute numerically, since it involves computing several entanglement entropies, and these are obtained using the replica trick, which means that one works with a system size that is twice as big as the physical system. In fact, for many systems, it is only possible to compute $\gamma$ for a range of system sizes that are too small to allow for an extrapolation to the thermodynamic limit. Here, we show that the transition point can be found by computing the charge of the anyons. This quantity can be expressed as a classical mean value and is much less expensive to compute. As a result, we can determine the transition point more accurately.

We consider a square lattice with a roughly circular boundary to mimic a quantum Hall droplet. The $N$ lattice sites are at the positions $z_1, \ldots, z_N$, and the local basis on site $j$ is $|n_j\rangle$, where $n_j \in \{0,1,2\}$ is the number of bosons on the site. The lattice Moore-Read state $|\Psi_0\rangle$ is defined as \cite{Nielsen2}
\begin{align}\label{wavefuncMR}
&|\Psi_x\rangle \propto \sum_{n_1,....,n_N}\Psi_x(n_1,\ldots,n_N) \ | n_1,...,n_N \rangle,\\
&\Psi_x(n_1,\ldots,n_N) =
\mathcal{G}^x_n \delta_n \prod_{i<j}(z_{i}-z_{j})^{2n_{i}n_{j}}	
\prod_{i\neq j}(z_{i}-z_{j})^{-\eta n_{i}},\nonumber
\end{align}
where $\mathcal{G}^0_n=\Pf[1/(z_i^\prime-z_j^\prime)]$, $\Pf(\ldots)$ is the Pfaffian, $z_i^\prime$ are the positions of the $\mathcal{M}$ singly occupied lattice sites, $\delta_n$ is a delta function that enforces the number of particles to be $M\equiv\sum_in_i=\eta N/2$, and $\eta$ is the magnetic flux per site. Note that we can vary the lattice filling factor $M/N=\eta/2$ by changing $\eta$.

We also introduce the states $|\Psi_a\rangle$ and $|\Psi_b\rangle$ with
\begin{align}
&\mathcal{G}^a_n =
2^{-\frac{\mathcal{M}}{2}}
\prod_{j}(w_2-{z_j})^{-n_j} \nonumber\\ &\times
\Pf\bigg[\frac{(z_i^\prime - w_1)(z_j^\prime - w_2)+(z_j^\prime - w_1)(z_i^\prime - w_2)}{(z_i^\prime-z_j^\prime)}\bigg], \label{wavefunc_I_tot}\\
&\mathcal{G}^b_n = \Pf\bigg(\frac{1}{z_i^\prime-z_j^\prime}\bigg)
\prod_{j}(w_1-z_j)^{n_j}\prod_{j}(w_2-z_j)^{-n_j}. \label{wavefunc_L_tot}
\end{align}
When the system is in the topological phase, the state $|\Psi_a\rangle$ ($|\Psi_b\rangle$) has an anyon of charge $+1/4$ ($+1/2$) at $w_1$ and an anyon of charge $-1/4$ ($-1/2$) at $w_2$ \cite{Nielsen1}. Few-body parent Hamiltonians for the states can be derived for a range of $\eta$ values and arbitrary choices of the lattice site positions $z_i$ \cite{Nielsen1,Nielsen2}.

\begin{figure}
\includegraphics[width=\columnwidth]{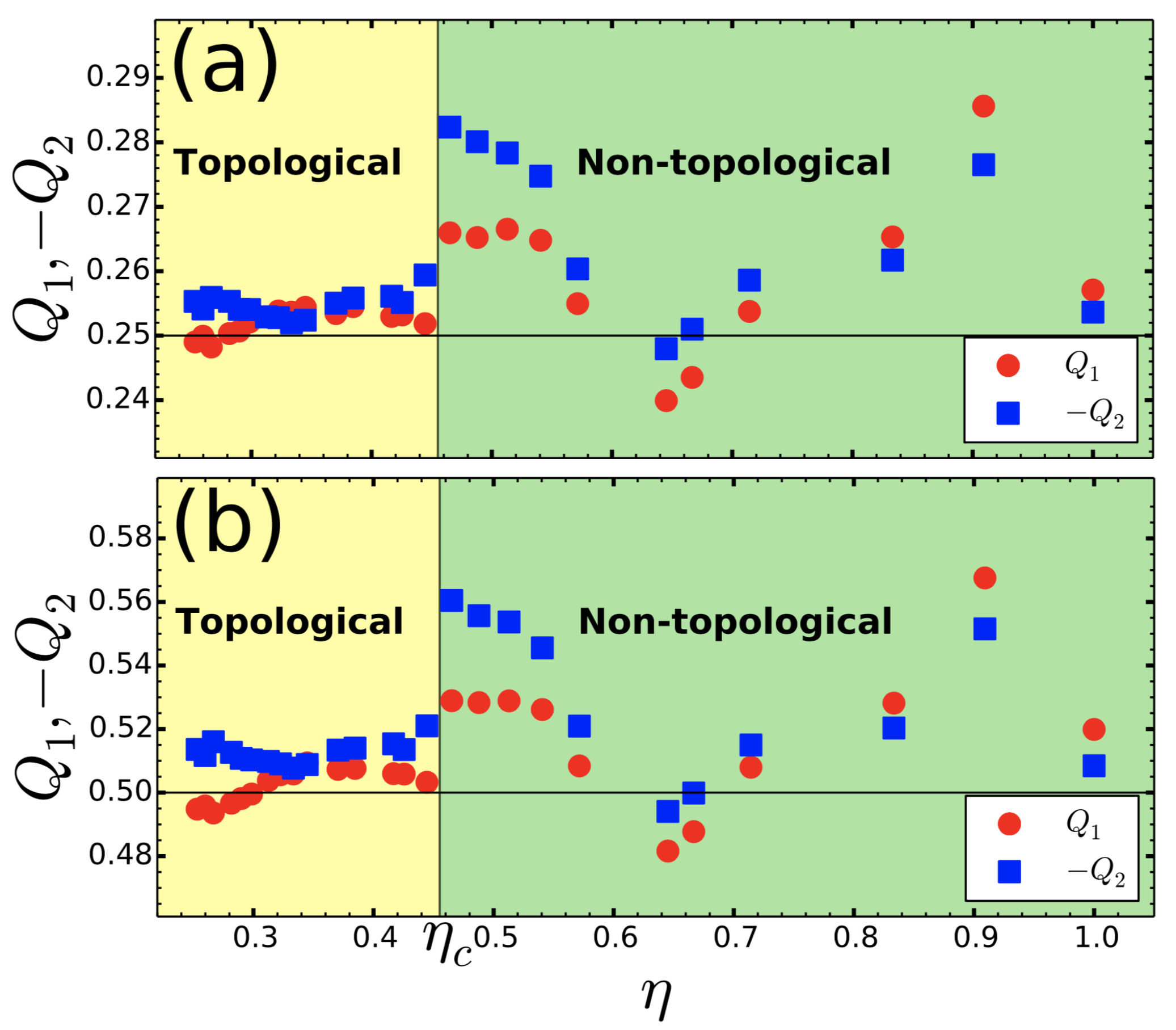}
\caption{(a) Excess charges $Q_1$ and $Q_2$ for the Moore-Read state $|\Psi_a\rangle$ on a square lattice as a function of the flux per site $\eta$. In the topological phase, $Q_1$ and $-Q_2$ are close to the charge of the positive anyon (horizontal line at $1/4$). In the nontopological phase, $Q_1$ and $Q_2$ may take any value. The jump away from $1/4$ predicts the transition point $\eta_c\in[0.44,0.46]$. (b) To test the robustness of the method, we observe that $|\Psi_b\rangle$ gives the same transition point. The Monte Carlo errors are of order $10^{-4}$.}\label{MR_TQPT}
\end{figure}

When anyons are present in the system, they modify the particle density in local regions around each $w_k$. Let us consider a circular region with radius $R$. If $R$ is large enough to enclose the anyon, but small enough to not enclose other anyons, the number of particles missing within the region, which is given by the excess charge
\begin{equation}\label{Excess_Charge}
\mathcal{Q}_k = -\sum_{i=1}^N \rho(z_i) \, \theta(R-|z_i-w_k|),
\end{equation}
equals the charge of the anyon at $w_k$. Here, $\theta(\ldots)$ is the Heaviside step function and
\begin{equation}\label{Density_Profile}
\rho(z_i) = \langle\Psi_{x} |n_i|\Psi_{x} \rangle
- \langle\Psi_{0} |n_i|\Psi_{0} \rangle, \quad x\in\{a,b\}
\end{equation}
is the density profile of the anyons. In the nontopological phase, a more complicated density pattern can arise. The expectation is hence that $\mathcal{Q}_k$ is close to the anyon charge in the topological phase and varies with $\eta$ in the nontopological phase, and we use this to detect the transition.

In Fig.\ \ref{MR_TQPT}, we choose $R=|w_1-w_2|/2$ and $M=40$, and we vary the number of lattice sites from $N=316$ to $N=80$ to achieve different $\eta$ values in the range $[1/4,1]$. We observe that the excess charges for $|\Psi_a\rangle$ are $Q_1\approx -Q_2\approx1/4$ for $\eta<\eta_c$ and fluctuate for $\eta>\eta_c$, where $\eta_c\in [0.44,0.46]$. As a test of the robustness of the approach, we observe that the same transition point is predicted using $|\Psi_b\rangle$. We have also checked that the fact that there is a jump in the excess charges from $\eta\simeq0.44$ to $\eta\simeq0.46$ is insensitive to the precise choice of the distance $|w_1-w_2|$.

\subsection{Moore-Read model on a fractal lattice}

We next consider the Moore-Read model on a fractal lattice. The fractal lattice is not periodic, and we can therefore not apply methods, such as ground state degeneracy, spectral flow, or many-body Chern number computations to detect a possible phase transition. Methods based on entanglement computations do also not apply, since we do not have a thorough understanding of entanglement properties of topological many-body states on fractal lattices. Fidelity cannot be used either, since the Hilbert space changes when the considered parameter changes. Quasiparticle properties, on the contrary, can detect a transition, as we will now show.

\begin{figure}
\includegraphics[width=\columnwidth]{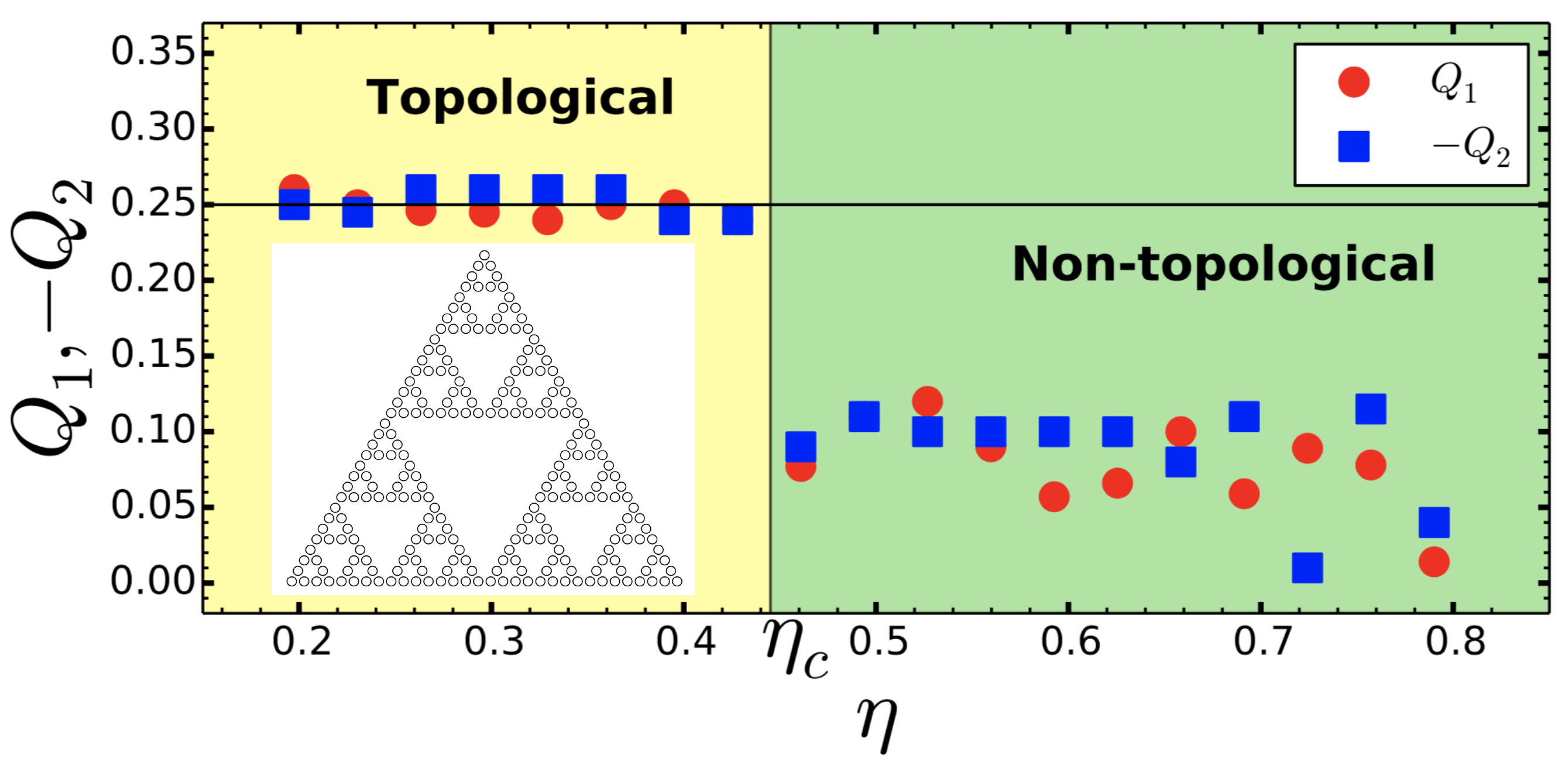}
\caption{Excess charges $Q_1$ and $Q_2$ for the Moore-Read state $|\Psi_a\rangle$ on a fractal lattice (inset) as a function of the flux per site $\eta$. In the topological phase, $Q_1$ and $-Q_2$ are close to the charge of the positive anyon (horizontal line at $1/4$). In the nontopological phase, $Q_1$ and $Q_2$ may take any value. The jump away from $1/4$ predicts the transition point $\eta_c\in[0.43,0.46]$. The Monte Carlo errors are of order $10^{-4}$.}\label{MR_TQPT_SG}
\end{figure}

Lattice Laughlin fractional quantum Hall models were recently constructed on fractals \cite{Manna1}, and we here consider a similar construction for the Moore-Read state. Specifically, we define the state $|\Psi_a\rangle$ on a lattice constructed from the Sierpi\'nski gasket with $N=243$ triangles by placing one lattice site on the center of each triangle. In Fig.\ \ref{MR_TQPT_SG}, we vary the particle number $M \in [24,96]$ to have different $\eta \in [0.19,0.79]$ values and plot the excess charges as a function of $\eta$. The excess charges are $Q_1\approx -Q_2\approx1/4$ for $\eta<\eta_c$ and fluctuate for $\eta>\eta_c$, which reveals a phase transition at the transition point $\eta_c\in [0.43,0.46]$.

\section{Disordered interacting Hofstadter model}

As another example, we study a Hofstadter model for hardcore bosons on a square lattice. The clean model is known to host a topological phase for low enough lattice filling factor \cite{Lukin1,Lukin2}, and here we investigate the effect of adding a disordered potential. The system sizes that can be reached with exact diagonalization are too small to allow for a computation of the topological entanglement entropy. Instead, we use the anyon charges to show that the system undergoes a phase transition as a function of the disorder strength. This gives a large speed up in computation time compared to the many-body Chern number computations that were done in \cite{Lukin1}. This is so, because it only takes two exact diagonalizations per date point to get the anyon charges, while the Chern number computation involves a large number of exact diagonalizations per data point, corresponding to a grid of twist angles in two dimensions.

The Hofstadter model describes particles hopping on a two-dimensional square lattice in the presence of a magnetic field perpendicular to the plane. Hopping is allowed between nearest neighbor sites, and the magnetic field is taken into account by making the hopping amplitudes complex. Whenever a particle hops around a closed loop, the wavefunction acquires a phase, which is equal to the Aharonov-Bohm phase for a charged particle encircling the same amount of magnetic flux.

\begin{figure}
\includegraphics[width=\columnwidth]{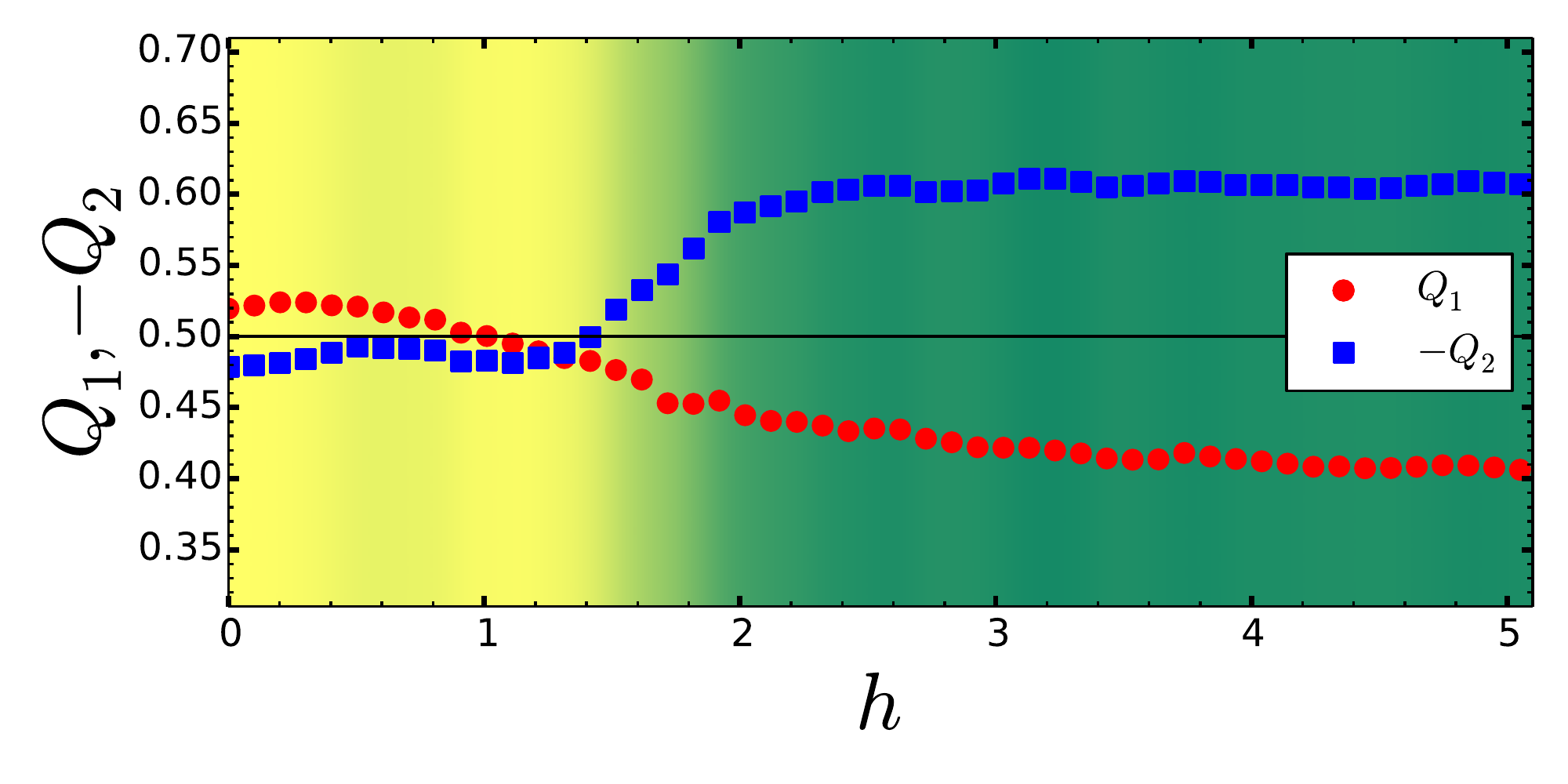}
\caption{Excess charges $Q_1$ and $Q_2$ as a function of the disorder strength $h$ for the interacting Hofstadter model with $M=3$, $N=24$, and $\alpha=0.25$. In the topological phase, $Q_1\approx -Q_2\approx 1/2$ (horizontal line), and the observed change away from this value predicts the transition point $h_c \simeq 1.5$. We average over $2000$ statistically independent disorder realizations for each $h$ to ensure convergence of the data.}\label{HCB_Dis_TQPT}
\end{figure}

We take open boundary conditions and add interactions by considering hardcore bosons. For a lattice with $N=L_x\times L_y$ sites, the Hamiltonian takes the form
\begin{eqnarray}\label{BH_1}
&& H_0 = - \sum_{x=1}^{L_x-1} \sum_{y=1}^{L_y} \left(  c^{\dagger}_{x+1,y} c_{x,y} e^{-i \pi \alpha y} + \text{H.c.} \right)\\
&&- \sum_{x=1}^{L_x} \sum_{y=1}^{L_y-1} \left(c^{\dagger}_{x,y+1} c_{x,y} e^{i \pi \alpha x} + \text{H.c.} \right)
+ \sum_{x=1}^{L_x}\sum_{y=1}^{L_y}h_{x,y} n_{x,y},\nonumber
\end{eqnarray}
where $c_{x,y}$ is the hardcore boson annihilation operator and $n_{x,y} = c^{\dagger}_{x,y} c_{x,y}$ is the number operator acting on the lattice site at the position $(x,y)$ with $x\in \{1,\ldots,L_x\}$ and $y\in \{1,\ldots,L_y\}$. If a particle hops around a plaquette, the phase acquired is $2\pi\alpha$, so $\alpha$ is the flux through the plaquette. We here consider the case, where the number of flux units per particle is two, i.e.\ $N\alpha/M=2$. The last term in \eqref{BH_1} is the disordered potential, and $h_{x,y} \in [-h,h]$ is drawn from a uniform distribution of width $2h$, where $h$ is the disorder strength.

In the clean model, it is well-known that one can trap anyons in the ground state by adding a local potential with a strength that is sufficiently large compared to the hopping amplitude \cite{kapitmueller,nielsen2018quasielectrons,eckardt}. We here choose
\begin{equation}\label{H_A}
H_V = V n_{x_1,y_1} - V n_{x_2,y_2}, \quad (x_1,y_1) \neq (x_2,y_2),
\end{equation}
where $V\gg 1$. This potential traps one positively (negatively) charged anyon at the site $(x_1,y_1)$ ($(x_2,y_2)$).

We use the excess charge in a region around the sites $(x_1,y_1)$ and $(x_2,y_2)$ to detect the phase transition. We define the density profile as
\begin{equation}\label{Density_Profile_1}
\rho(x+iy)=\langle n_{x,y} \rangle_{H_0 + H_V} - \langle n_{x,y} \rangle_{H_0},
\end{equation}
where $\langle n_{x,y} \rangle_{H_0 + H_V}$ is the particle density, when the trapping potential is present, and $\langle n_{x,y} \rangle_{H_0}$ is the particle density, when the trapping potential is absent. The excess charge is then defined as in \eqref{Excess_Charge} with $w_1=x_1+iy_1$ and $w_2=x_2+iy_2$. Here, we choose $R$ such that the circular region includes all sites up to the second nearest neighbor sites. The absolute value of the excess charge should be close to $1/2$ in the topological region, while it can take any value and may vary with $h$ in the nontopological region.

We choose a point, which is deep in the topological phase for $h=0$, namely $M=3$, $N=24$, and $\alpha=0.25$, and plot the excess charges as a function of the disorder strength $h$ in Fig.\ \ref{HCB_Dis_TQPT}. We observe that $Q_1$ and $-Q_2$ are close to $1/2$ up to $h \simeq 1.5$, while the excess charges deviate more from $1/2$ for $h > 1.5$. The data hence predict the phase transition to happen at $h_c\simeq 1.5$.

\begin{figure}
\includegraphics[width=\columnwidth]{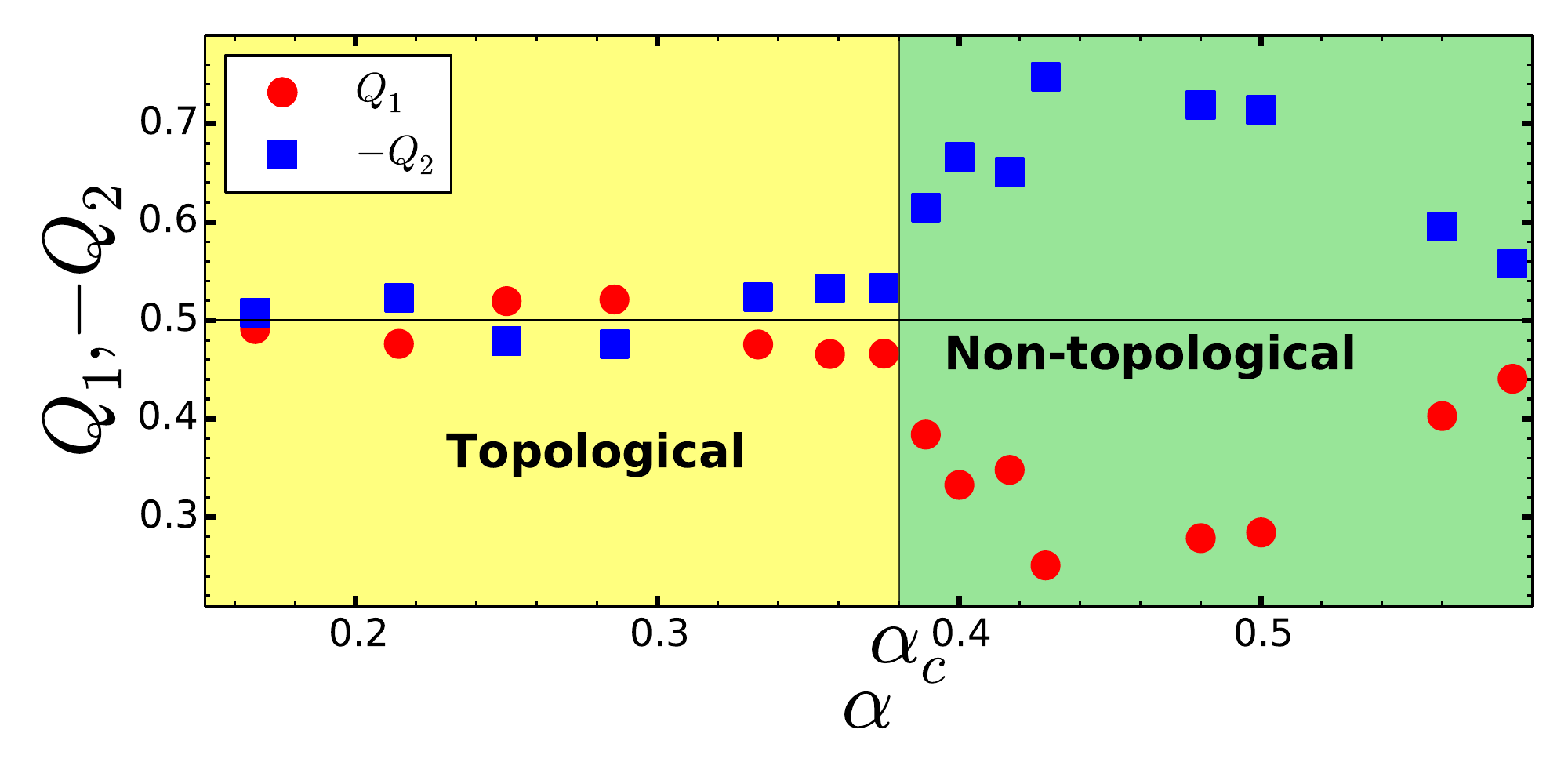}
\caption{Excess charges $Q_1$ and $Q_2$ as a function of the magnetic flux per plaquette $\alpha$ for the interacting Hofstadter model without disorder. In the topological phase, $Q_1\approx -Q_2\approx 1/2$, and the observed change away from this value predicts the transition point $\alpha_c\in[0.375,0.389]$.}\label{HCB_TQPT}
\end{figure}

We can also put the disorder strength to $h=0$ and study the clean model as a function of the magnetic flux $\alpha$. Specifically, we vary $\alpha$ and the lattice filling $M/N$, while keeping the flux per particle $N\alpha/M$ fixed. For this case, it was found in \cite{Lukin1} that there is a phase transition at $\alpha_c\in[0.375,0.400]$. We take values of $M$ and $N$ (see Tab.\ \ref{HCB_data}), which are numerically accessible for exact diagonalization, and for each choice $\alpha = 2M/N$. Figure \ref{HCB_TQPT} shows that $Q_1$ and $-Q_2$ are quite close to $1/2$ for $\alpha$ values up to $0.375$, but for higher $\alpha$ they deviate much more from $1/2$. The data hence predict the transition point $\alpha_c\in[0.375,0.389]$, which is consistent with the result in \cite{Lukin1}.

\begin{table}
\begin{tabular}{|c|c|c|c|c|c|c|c|c|}
\hline
{\em $M$} & $N$ & {$ L_x \times L_y$ } & {\em $\alpha$}  &
{\text{dim}$(\mathcal{H})$} & {\em $Q_{+}$}& {\em $ Q_{-}$}  \\
\hline
\hline 2 & 24 & 6 $\times$ 4 & 0.167 &     276 & 0.491 & 0.507 \\
\hline 3 & 28 & 7 $\times$ 4 & 0.214 &    3276 & 0.476 & 0.522 \\
\hline 3 & 24 & 6 $\times$ 4 & 0.250 &    2024 & 0.519 & 0.478 \\
\hline 4 & 28 & 7 $\times$ 4 & 0.286 &   20475 & 0.521 & 0.475 \\
\hline 4 & 24 & 6 $\times$ 4 & 0.333 &   10626 & 0.475 & 0.520 \\
\hline 5 & 28 & 7 $\times$ 4 & 0.357 &   98280 & 0.465 & 0.532 \\
\hline 6 & 32 & 8 $\times$ 4 & 0.375 &  906192 & 0.466 & 0.533 \\
\hline 7 & 36 & 6 $\times$ 6 & 0.389 & 8347680 & 0.380 & 0.610 \\
\hline 5 & 25 & 5 $\times$ 5 & 0.400 & 	 53130 & 0.332 & 0.666 \\
\hline 5 & 24 & 6 $\times$ 4 & 0.417 &   42504 & 0.348 & 0.650 \\
\hline 6 & 28 & 7 $\times$ 4 & 0.429 &  376740 & 0.250 & 0.747 \\
\hline 6 & 25 & 5 $\times$ 5 & 0.480 &  177100 & 0.278 & 0.719 \\
\hline 7 & 28 & 7 $\times$ 4 & 0.500 & 1184040 & 0.284 & 0.714 \\
\hline 7 & 25 & 5 $\times$ 5 & 0.560 &  480700 & 0.402 & 0.595 \\
\hline 7 & 24 & 6 $\times$ 4 & 0.583 &  346104 & 0.440 & 0.557 \\
\hline
\end{tabular}
\caption{We show here the different choices we make for the number of particles $M$, the shapes and sizes $N=L_x \times L_y$ of the lattices, and the fluxes per plaquette $\alpha = 2M/N$. The quantity dim$(\mathcal{H})$ is the dimension of the corresponding Hilbert spaces. We display the data for the absolute values of the excess charges $Q_{+}$ and $Q_{-}$. There is a significant change in $Q_{+}$ and $Q_{-}$, when going from $\alpha = 0.375$ to $\alpha \simeq 0.389$.}\label{HCB_data}
\end{table}

\section{Toric code in a magnetic field}

To test the applicability of the method outside the family of chiral fractional quantum Hall models, we next study Kitaev's toric code \cite{Kitaev1,Kitaev2} on a square lattice with periodic boundary conditions. This system exhibits a $\mathbb{Z}_2$ topologically ordered phase, and it is known that a sufficiently strong, uniform magnetic field drives the system into a polarized phase \cite{Vidal4,Vidal2,Vidal3,zarei,greplova}. Here we show that anyons inserted into the system are able to detect this phase transition, and the obtained transition points agree with earlier results based on perturbative, analytical calculations and tensor network studies. Our computations rely on exact diagonalization for a system with $18$ spins and are hence quite fast to do numerically. We find that the anyons are significantly better at predicting the phase transition point than the energy gap closing for the same system size.

The toric code has a spin-$1/2$ on each of the edges of the $N_x\times N_y$ square lattice. The Hamiltonian
\begin{equation}\label{H_TC1}
H_{\text{TC}} = - \sum_{p} B_p - \sum_{v} A_v, \; B_p = \prod_{i \in p} \sigma^z_i, \; A_v = \prod_{i \in v} \sigma^x_i,
\end{equation}
is expressed in terms of the Pauli operators $\sigma^x_i$ and $\sigma^z_i$, which act on the $N=2N_xN_y$ spins. The sums are over all plaquettes $p$ and vertices $v$ of the lattice. $B_p$ acts on the spins on the four edges surrounding the plaquette $p$, and $A_v$ acts on the spins on the four edges connecting to the vertex $v$.

$H_{\text{TC}}$ is exactly solvable, and the four degenerate ground states are eigenstates of $B_p$ and $A_v$ with eigenvalue $1$. States containing anyons are obtained by applying certain string operators to the ground states. The string operator either changes the eigenvalue of two $A_v$ operators to $-1$ or the eigenvalue of two $B_p$ operators to $-1$. In the former case, two electric excitations $e_v$ are created, and in the latter case two magnetic excitations $m_p$ are created. The wavefunction acquires a minus sign if one $m_p$ is moved around one $e_v$, and the excitations are hence Abelian anyons.

Here, we instead modify the Hamiltonian, such that anyons are present in the ground states. The ground states of the Hamiltonian $H_m\equiv H_{\text{TC}} + 2 B_{p_1} + 2 B_{p_2}$ have one $m_p$ on each of the plaquettes $p_1$ and $p_2$. Similarly, the ground states of the Hamiltonian $H_e \equiv H_{\text{TC}} + 2A_{v_1}+ 2A_{v_2}$ have one $e_v$ on each of the vertices $v_1$ and $v_2$. We drive the system through a phase transition by adding a magnetic field $H_\lambda^k = \lambda \sum_{i} \sigma^k_i$ in the $k$-direction with strength $\lambda$. When $\lambda$ is large enough, it is energetically favorable to polarize all the spins, and the system is no longer topological.

\begin{figure}
\includegraphics[width=\columnwidth]{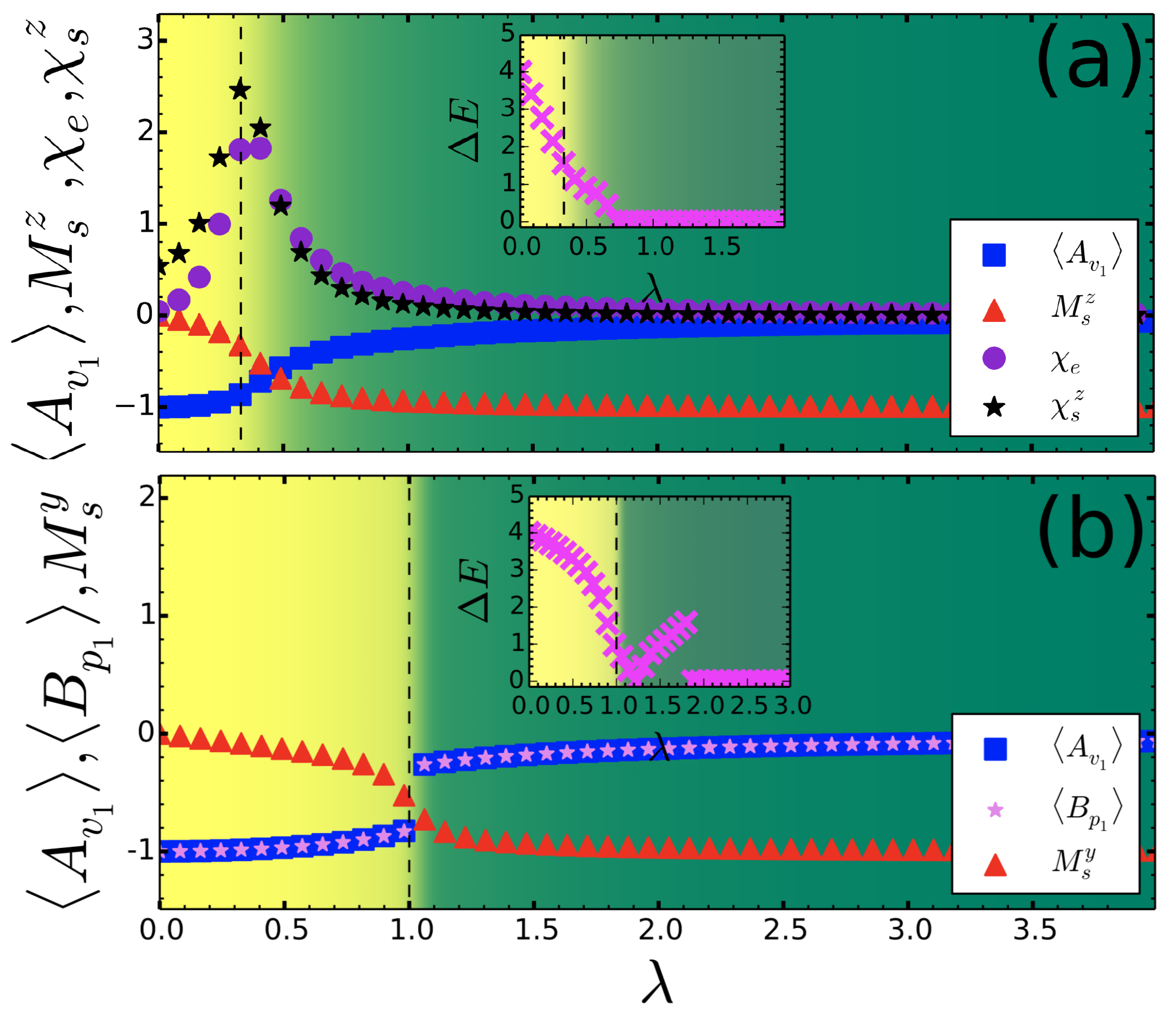}
\caption{(a) The toric code with a magnetic field of strength $\lambda$ in the $z$ direction undergoes a phase transition at $\lambda_c \simeq 0.33$. The transition is seen in $\langle  A_{v_1} \rangle$, which detects the anyons in the ground states of $H_e+H_\lambda^z$, and in $M_s^z=\frac{1}{N} \langle \sum_{i} \sigma^k_i \rangle$, which is the magnetization per spin for the ground states of $H_{\textrm{TC}}+H_\lambda^z$. We also show $\chi_e=\partial \langle  A_{v_1} \rangle/\partial \lambda$ and $\chi_s^z=-\partial M_s^z/\partial \lambda$. (b) When the magnetic field is in the $y$ direction, the phase transition happens at $\lambda_c = 1$. We plot $\langle  A_{v_1} \rangle$ for $H_e+H_\lambda^y$ and $\langle  B_{p_1} \rangle$ for $H_m+H_\lambda^y$ as well as the magnetization per spin $M_s^y=\frac{1}{N} \langle \sum_{i} \sigma^y_i \rangle$ for $H_{\textrm{TC}}+H_\lambda^y$. The closing of the energy gap $\Delta E$ between the ground state (fourfold degenerate) and the first excited state of $H_{\textrm{TC}}+H_\lambda^k$ shown in the insets does not accurately predict the transition points. We use a $3\times3$ lattice with $18$ spins in all cases, and the color gradient from yellow (topological phase) to green (nontopological phase) is plotted according to the value of $\langle  A_{v_1} \rangle$.} \label{TC}
\end{figure}

Previous investigations, based on perturbative, analytical calculations and tensor network studies \cite{Vidal4,Vidal2,Vidal3}, have shown that $H_{\text{TC}}+H_\lambda^z$ has a second order phase transition at $\lambda_c\simeq 0.33$, while $H_{\text{TC}}+H_\lambda^y$ has a first order phase transition at $\lambda_c=1$. The magnetization per spin computed using exact diagonalization (Fig.\ \ref{TC}) gives similar values for the transition points.

We now use anyons to detect the transition. We study $\langle  A_{v_1} \rangle=\langle  A_{v_2} \rangle$ for the Hamiltonian $H_e+H_\lambda^k$ and $\langle  B_{p_1} \rangle=\langle  B_{p_2} \rangle$ for the Hamiltonian $H_m+H_\lambda^k$. $\langle  A_{v_1} \rangle=-1$ and $\langle  B_{p_1} \rangle=-1$ signify the presence of the anyons. In the fully polarized phase, both $\langle  A_{v_1} \rangle$ and $\langle  B_{p_1} \rangle$ vanish when $k=y$ and $\langle  A_{v_1} \rangle$ vanishes and $\langle  B_{p_1} \rangle\to +1$ when $k=z$.

The transition seen in $\langle  A_{v_1} \rangle$ for the ground states of $H_e+H_\lambda^z$ in Fig.\ \ref{TC}(a) is consistent with $\lambda_c \simeq 0.33$. Both the transition point and the width of the transition region, which is due to finite size effects, are comparable to the same quantities obtained from the magnetic order parameter. The anyons predict the transition point more accurately than the energy gap closing, which, for the same system size, happens only at $\lambda\simeq0.7$.

$\langle  B_{p_1} \rangle$ is not suitable for detecting the transition because $B_{p_1}$ commutes with all terms in $H_m+H_\lambda^z$. All energy eigenstates are therefore also eigenstates of $B_{p_1}$ with eigenvalue $\pm1$. As a result, $\langle  B_{p_1} \rangle$ only measures whether the ground states have $B_{p_1}$ eigenvalue $+1$ or $-1$. The first transition to a ground state with $B_{p_1}$ eigenvalue $+1$ happens around $\lambda \simeq 2.08$, but this does not exclude gap closings at smaller $\lambda$ values. These problems do not occur for $\langle  A_{v_1} \rangle$, since $A_{v_1}$ does not commute with $H_e+H_\lambda^z$.

The transition seen in $\langle  A_{v_1} \rangle$ ($\langle  B_{p_1} \rangle$) in Fig.\ \ref{TC}(b) for the ground states of $H_e+H_\lambda^y$ ($H_m+H_\lambda^y$) is consistent with $\lambda_c = 1$, and the transition is sharper than for the magnetic order parameter. The anyons also better predict the transition point than the energy gap closing, which happens around $\lambda\simeq 1.2$ for the same system size.

\section{Conclusions}

We have shown that properties of quasiparticles are an interesting tool to detect topological quantum phase transitions. The approach is to trap anyons in the ground state and study how their properties change, when the system crosses a phase transition. If we are able to create quasiparticles with robust and nontrivial braiding properties in a system, we know that the system hosts anyons of the observed type. For several quite different examples, we have demonstrated, however, that the phase transitions can be detected by observing a simple property, such as the charge of the anyons. This means that the phase transition points can be computed at low numerical costs.

The approach suggested here to detect topological quantum phase transitions is particularly direct, since, to fully exploit the interesting properties of topologically ordered systems, one needs to be able to create anyons in the systems and detect their properties. In the interacting Hofstadter model, the anyons can be created by adding a local potential, and the charge of the anyons used to detect the phase transition can be measured by measuring the expectation value of the number of particles on each site. Both of these can be done in experiments with ultracold atoms in optical lattices \cite{weitenberg,eckardt}. For the toric code model, the complexity of generating the Hamiltonians including the terms creating the anyons is about the same as generating the Hamiltonian without these terms, and first steps towards realizing the Hamiltonian in experiments have been taken \cite{Toric1,Toric2,Toric3}.

The ideas presented in this work can be applied as long as the two phases support quasiparticles with different properties and one can find suitable ways to create the quasiparticles. It is well-suited for detecting phase transitions between different topologically ordered phases. In addition, it would be interesting to investigate what we can learn about transitions between nontopological phases by studying quasiparticles.

\begin{acknowledgements}
We thank Julien Vidal for helpful comments regarding previous work on phase transitions in the toric code in a magnetic field. J.W. acknowledges NSF Grant No.\ DMR $1306897$ and NSF Grant No.\ DMR $1056536$ for partial support.
\end{acknowledgements}

\bibliography{bibfile}

\begin{thebibliography}{43}%
\makeatletter
\providecommand \@ifxundefined [1]{%
 \@ifx{#1\undefined}
}%
\providecommand \@ifnum [1]{%
 \ifnum #1\expandafter \@firstoftwo
 \else \expandafter \@secondoftwo
 \fi
}%
\providecommand \@ifx [1]{%
 \ifx #1\expandafter \@firstoftwo
 \else \expandafter \@secondoftwo
 \fi
}%
\providecommand \natexlab [1]{#1}%
\providecommand \enquote  [1]{``#1''}%
\providecommand \bibnamefont  [1]{#1}%
\providecommand \bibfnamefont [1]{#1}%
\providecommand \citenamefont [1]{#1}%
\providecommand \href@noop [0]{\@secondoftwo}%
\providecommand \href [0]{\begingroup \@sanitize@url \@href}%
\providecommand \@href[1]{\@@startlink{#1}\@@href}%
\providecommand \@@href[1]{\endgroup#1\@@endlink}%
\providecommand \@sanitize@url [0]{\catcode `\\12\catcode `\$12\catcode
  `\&12\catcode `\#12\catcode `\^12\catcode `\_12\catcode `\%12\relax}%
\providecommand \@@startlink[1]{}%
\providecommand \@@endlink[0]{}%
\providecommand \url  [0]{\begingroup\@sanitize@url \@url }%
\providecommand \@url [1]{\endgroup\@href {#1}{\urlprefix }}%
\providecommand \urlprefix  [0]{URL }%
\providecommand \Eprint [0]{\href }%
\providecommand \doibase [0]{http://dx.doi.org/}%
\providecommand \selectlanguage [0]{\@gobble}%
\providecommand \bibinfo  [0]{\@secondoftwo}%
\providecommand \bibfield  [0]{\@secondoftwo}%
\providecommand \translation [1]{[#1]}%
\providecommand \BibitemOpen [0]{}%
\providecommand \bibitemStop [0]{}%
\providecommand \bibitemNoStop [0]{.\EOS\space}%
\providecommand \EOS [0]{\spacefactor3000\relax}%
\providecommand \BibitemShut  [1]{\csname bibitem#1\endcsname}%
\let\auto@bib@innerbib\@empty
\bibitem [{\citenamefont {Sachdev}(2011)}]{Sachdev1}%
  \BibitemOpen
  \bibfield  {author} {\bibinfo {author} {\bibfnamefont {S.}~\bibnamefont
  {Sachdev}},\ }\href@noop {} {\emph {\bibinfo {title} {Quantum Phase
  Transitions}}}\ (\bibinfo  {publisher} {Cambridge University Press,
  Cambridge, England},\ \bibinfo {year} {2011})\BibitemShut {NoStop}%
\bibitem [{\citenamefont {Wen}(2017)}]{Wen5}%
  \BibitemOpen
  \bibfield  {author} {\bibinfo {author} {\bibfnamefont {X.-G.}\ \bibnamefont
  {Wen}},\ }\href {\doibase 10.1103/RevModPhys.89.041004} {\bibfield  {journal}
  {\bibinfo  {journal} {Rev. Mod. Phys.}\ }\textbf {\bibinfo {volume} {89}},\
  \bibinfo {pages} {041004} (\bibinfo {year} {2017})}\BibitemShut {NoStop}%
\bibitem [{\citenamefont {Grushin}\ \emph {et~al.}(2015)\citenamefont
  {Grushin}, \citenamefont {Motruk}, \citenamefont {Zaletel},\ and\
  \citenamefont {Pollmann}}]{Pollmann1}%
  \BibitemOpen
  \bibfield  {author} {\bibinfo {author} {\bibfnamefont {A.~G.}\ \bibnamefont
  {Grushin}}, \bibinfo {author} {\bibfnamefont {J.}~\bibnamefont {Motruk}},
  \bibinfo {author} {\bibfnamefont {M.~P.}\ \bibnamefont {Zaletel}}, \ and\
  \bibinfo {author} {\bibfnamefont {F.}~\bibnamefont {Pollmann}},\ }\href
  {\doibase 10.1103/PhysRevB.91.035136} {\bibfield  {journal} {\bibinfo
  {journal} {Phys. Rev. B}\ }\textbf {\bibinfo {volume} {91}},\ \bibinfo
  {pages} {035136} (\bibinfo {year} {2015})}\BibitemShut {NoStop}%
\bibitem [{\citenamefont {Wildeboer}\ \emph {et~al.}(2017)\citenamefont
  {Wildeboer}, \citenamefont {Seidel},\ and\ \citenamefont {Melko}}]{Julia1}%
  \BibitemOpen
  \bibfield  {author} {\bibinfo {author} {\bibfnamefont {J.}~\bibnamefont
  {Wildeboer}}, \bibinfo {author} {\bibfnamefont {A.}~\bibnamefont {Seidel}}, \
  and\ \bibinfo {author} {\bibfnamefont {R.~G.}\ \bibnamefont {Melko}},\ }\href
  {\doibase 10.1103/PhysRevB.95.100402} {\bibfield  {journal} {\bibinfo
  {journal} {Phys. Rev. B}\ }\textbf {\bibinfo {volume} {95}},\ \bibinfo
  {pages} {100402(R)} (\bibinfo {year} {2017})}\BibitemShut {NoStop}%
\bibitem [{\citenamefont {Wen}\ and\ \citenamefont {Niu}(1990)}]{WenNiu}%
  \BibitemOpen
  \bibfield  {author} {\bibinfo {author} {\bibfnamefont {X.~G.}\ \bibnamefont
  {Wen}}\ and\ \bibinfo {author} {\bibfnamefont {Q.}~\bibnamefont {Niu}},\
  }\href {\doibase 10.1103/PhysRevB.41.9377} {\bibfield  {journal} {\bibinfo
  {journal} {Phys. Rev. B}\ }\textbf {\bibinfo {volume} {41}},\ \bibinfo
  {pages} {9377} (\bibinfo {year} {1990})}\BibitemShut {NoStop}%
\bibitem [{\citenamefont {Tao}\ and\ \citenamefont {Haldane}(1986)}]{Haldane1}%
  \BibitemOpen
  \bibfield  {author} {\bibinfo {author} {\bibfnamefont {R.}~\bibnamefont
  {Tao}}\ and\ \bibinfo {author} {\bibfnamefont {F.~D.~M.}\ \bibnamefont
  {Haldane}},\ }\href {\doibase 10.1103/PhysRevB.33.3844} {\bibfield  {journal}
  {\bibinfo  {journal} {Phys. Rev. B}\ }\textbf {\bibinfo {volume} {33}},\
  \bibinfo {pages} {3844} (\bibinfo {year} {1986})}\BibitemShut {NoStop}%
\bibitem [{\citenamefont {Niu}\ \emph {et~al.}(1985)\citenamefont {Niu},
  \citenamefont {Thouless},\ and\ \citenamefont {Wu}}]{Wu1}%
  \BibitemOpen
  \bibfield  {author} {\bibinfo {author} {\bibfnamefont {Q.}~\bibnamefont
  {Niu}}, \bibinfo {author} {\bibfnamefont {D.~J.}\ \bibnamefont {Thouless}}, \
  and\ \bibinfo {author} {\bibfnamefont {Y.-S.}\ \bibnamefont {Wu}},\ }\href
  {\doibase 10.1103/PhysRevB.31.3372} {\bibfield  {journal} {\bibinfo
  {journal} {Phys. Rev. B}\ }\textbf {\bibinfo {volume} {31}},\ \bibinfo
  {pages} {3372} (\bibinfo {year} {1985})}\BibitemShut {NoStop}%
\bibitem [{\citenamefont {Kudo}\ \emph {et~al.}(2019)\citenamefont {Kudo},
  \citenamefont {Watanabe}, \citenamefont {Kariyado},\ and\ \citenamefont
  {Hatsugai}}]{Hatsugai1}%
  \BibitemOpen
  \bibfield  {author} {\bibinfo {author} {\bibfnamefont {K.}~\bibnamefont
  {Kudo}}, \bibinfo {author} {\bibfnamefont {H.}~\bibnamefont {Watanabe}},
  \bibinfo {author} {\bibfnamefont {T.}~\bibnamefont {Kariyado}}, \ and\
  \bibinfo {author} {\bibfnamefont {Y.}~\bibnamefont {Hatsugai}},\ }\href
  {\doibase 10.1103/PhysRevLett.122.146601} {\bibfield  {journal} {\bibinfo
  {journal} {Phys. Rev. Lett.}\ }\textbf {\bibinfo {volume} {122}},\ \bibinfo
  {pages} {146601} (\bibinfo {year} {2019})}\BibitemShut {NoStop}%
\bibitem [{\citenamefont {Neupert}\ \emph {et~al.}(2011)\citenamefont
  {Neupert}, \citenamefont {Santos}, \citenamefont {Chamon},\ and\
  \citenamefont {Mudry}}]{Mudry1}%
  \BibitemOpen
  \bibfield  {author} {\bibinfo {author} {\bibfnamefont {T.}~\bibnamefont
  {Neupert}}, \bibinfo {author} {\bibfnamefont {L.}~\bibnamefont {Santos}},
  \bibinfo {author} {\bibfnamefont {C.}~\bibnamefont {Chamon}}, \ and\ \bibinfo
  {author} {\bibfnamefont {C.}~\bibnamefont {Mudry}},\ }\href {\doibase
  10.1103/PhysRevLett.106.236804} {\bibfield  {journal} {\bibinfo  {journal}
  {Phys. Rev. Lett.}\ }\textbf {\bibinfo {volume} {106}},\ \bibinfo {pages}
  {236804} (\bibinfo {year} {2011})}\BibitemShut {NoStop}%
\bibitem [{\citenamefont {Hu}\ \emph {et~al.}(2015)\citenamefont {Hu},
  \citenamefont {Gong}, \citenamefont {Zhu},\ and\ \citenamefont
  {Sheng}}]{Sheng1}%
  \BibitemOpen
  \bibfield  {author} {\bibinfo {author} {\bibfnamefont {W.-J.}\ \bibnamefont
  {Hu}}, \bibinfo {author} {\bibfnamefont {S.-S.}\ \bibnamefont {Gong}},
  \bibinfo {author} {\bibfnamefont {W.}~\bibnamefont {Zhu}}, \ and\ \bibinfo
  {author} {\bibfnamefont {D.~N.}\ \bibnamefont {Sheng}},\ }\href {\doibase
  10.1103/PhysRevB.92.140403} {\bibfield  {journal} {\bibinfo  {journal} {Phys.
  Rev. B}\ }\textbf {\bibinfo {volume} {92}},\ \bibinfo {pages} {140403(R)}
  (\bibinfo {year} {2015})}\BibitemShut {NoStop}%
\bibitem [{\citenamefont {Regnault}\ and\ \citenamefont
  {Bernevig}(2011)}]{Regnault}%
  \BibitemOpen
  \bibfield  {author} {\bibinfo {author} {\bibfnamefont {N.}~\bibnamefont
  {Regnault}}\ and\ \bibinfo {author} {\bibfnamefont {B.~A.}\ \bibnamefont
  {Bernevig}},\ }\href {\doibase 10.1103/PhysRevX.1.021014} {\bibfield
  {journal} {\bibinfo  {journal} {Phys. Rev. X}\ }\textbf {\bibinfo {volume}
  {1}},\ \bibinfo {pages} {021014} (\bibinfo {year} {2011})}\BibitemShut
  {NoStop}%
\bibitem [{\citenamefont {Thomale}\ \emph {et~al.}(2010)\citenamefont
  {Thomale}, \citenamefont {Sterdyniak}, \citenamefont {Regnault},\ and\
  \citenamefont {Bernevig}}]{Thomale1}%
  \BibitemOpen
  \bibfield  {author} {\bibinfo {author} {\bibfnamefont {R.}~\bibnamefont
  {Thomale}}, \bibinfo {author} {\bibfnamefont {A.}~\bibnamefont {Sterdyniak}},
  \bibinfo {author} {\bibfnamefont {N.}~\bibnamefont {Regnault}}, \ and\
  \bibinfo {author} {\bibfnamefont {B.~A.}\ \bibnamefont {Bernevig}},\ }\href
  {\doibase 10.1103/PhysRevLett.104.180502} {\bibfield  {journal} {\bibinfo
  {journal} {Phys. Rev. Lett.}\ }\textbf {\bibinfo {volume} {104}},\ \bibinfo
  {pages} {180502} (\bibinfo {year} {2010})}\BibitemShut {NoStop}%
\bibitem [{\citenamefont {Hermanns}\ \emph {et~al.}(2011)\citenamefont
  {Hermanns}, \citenamefont {Chandran}, \citenamefont {Regnault},\ and\
  \citenamefont {Bernevig}}]{Hermanns1}%
  \BibitemOpen
  \bibfield  {author} {\bibinfo {author} {\bibfnamefont {M.}~\bibnamefont
  {Hermanns}}, \bibinfo {author} {\bibfnamefont {A.}~\bibnamefont {Chandran}},
  \bibinfo {author} {\bibfnamefont {N.}~\bibnamefont {Regnault}}, \ and\
  \bibinfo {author} {\bibfnamefont {B.~A.}\ \bibnamefont {Bernevig}},\ }\href
  {\doibase 10.1103/PhysRevB.84.121309} {\bibfield  {journal} {\bibinfo
  {journal} {Phys. Rev. B}\ }\textbf {\bibinfo {volume} {84}},\ \bibinfo
  {pages} {121309(R)} (\bibinfo {year} {2011})}\BibitemShut {NoStop}%
\bibitem [{\citenamefont {Sterdyniak}\ \emph {et~al.}(2011)\citenamefont
  {Sterdyniak}, \citenamefont {Regnault},\ and\ \citenamefont
  {Bernevig}}]{Sterdyniak1}%
  \BibitemOpen
  \bibfield  {author} {\bibinfo {author} {\bibfnamefont {A.}~\bibnamefont
  {Sterdyniak}}, \bibinfo {author} {\bibfnamefont {N.}~\bibnamefont
  {Regnault}}, \ and\ \bibinfo {author} {\bibfnamefont {B.~A.}\ \bibnamefont
  {Bernevig}},\ }\href {\doibase 10.1103/PhysRevLett.106.100405} {\bibfield
  {journal} {\bibinfo  {journal} {Phys. Rev. Lett.}\ }\textbf {\bibinfo
  {volume} {106}},\ \bibinfo {pages} {100405} (\bibinfo {year}
  {2011})}\BibitemShut {NoStop}%
\bibitem [{\citenamefont {Levin}\ and\ \citenamefont {Wen}(2006)}]{Wen1}%
  \BibitemOpen
  \bibfield  {author} {\bibinfo {author} {\bibfnamefont {M.}~\bibnamefont
  {Levin}}\ and\ \bibinfo {author} {\bibfnamefont {X.-G.}\ \bibnamefont
  {Wen}},\ }\href {\doibase 10.1103/PhysRevLett.96.110405} {\bibfield
  {journal} {\bibinfo  {journal} {Phys. Rev. Lett.}\ }\textbf {\bibinfo
  {volume} {96}},\ \bibinfo {pages} {110405} (\bibinfo {year}
  {2006})}\BibitemShut {NoStop}%
\bibitem [{\citenamefont {Kitaev}\ and\ \citenamefont
  {Preskill}(2006)}]{Preskill1}%
  \BibitemOpen
  \bibfield  {author} {\bibinfo {author} {\bibfnamefont {A.}~\bibnamefont
  {Kitaev}}\ and\ \bibinfo {author} {\bibfnamefont {J.}~\bibnamefont
  {Preskill}},\ }\href {\doibase 10.1103/PhysRevLett.96.110404} {\bibfield
  {journal} {\bibinfo  {journal} {Phys. Rev. Lett.}\ }\textbf {\bibinfo
  {volume} {96}},\ \bibinfo {pages} {110404} (\bibinfo {year}
  {2006})}\BibitemShut {NoStop}%
\bibitem [{\citenamefont {Jiang}\ \emph {et~al.}(2012)\citenamefont {Jiang},
  \citenamefont {Wang},\ and\ \citenamefont {Balents}}]{Balents1}%
  \BibitemOpen
  \bibfield  {author} {\bibinfo {author} {\bibfnamefont {H.-C.}\ \bibnamefont
  {Jiang}}, \bibinfo {author} {\bibfnamefont {Z.}~\bibnamefont {Wang}}, \ and\
  \bibinfo {author} {\bibfnamefont {L.}~\bibnamefont {Balents}},\ }\href
  {https://doi.org/10.1038/nphys2465} {\bibfield  {journal} {\bibinfo
  {journal} {Nature Physics}\ }\textbf {\bibinfo {volume} {8}},\ \bibinfo
  {pages} {902–905} (\bibinfo {year} {2012})}\BibitemShut {NoStop}%
\bibitem [{\citenamefont {Zanardi}\ and\ \citenamefont
  {Paunkovi\ifmmode~\acute{c}\else \'{c}\fi{}}(2006)}]{zanardi}%
  \BibitemOpen
  \bibfield  {author} {\bibinfo {author} {\bibfnamefont {P.}~\bibnamefont
  {Zanardi}}\ and\ \bibinfo {author} {\bibfnamefont {N.}~\bibnamefont
  {Paunkovi\ifmmode~\acute{c}\else \'{c}\fi{}}},\ }\href {\doibase
  10.1103/PhysRevE.74.031123} {\bibfield  {journal} {\bibinfo  {journal} {Phys.
  Rev. E}\ }\textbf {\bibinfo {volume} {74}},\ \bibinfo {pages} {031123}
  (\bibinfo {year} {2006})}\BibitemShut {NoStop}%
\bibitem [{\citenamefont {Arovas}\ \emph {et~al.}(1984)\citenamefont {Arovas},
  \citenamefont {Schrieffer},\ and\ \citenamefont {Wilczek}}]{AroSchWil}%
  \BibitemOpen
  \bibfield  {author} {\bibinfo {author} {\bibfnamefont {D.}~\bibnamefont
  {Arovas}}, \bibinfo {author} {\bibfnamefont {J.~R.}\ \bibnamefont
  {Schrieffer}}, \ and\ \bibinfo {author} {\bibfnamefont {F.}~\bibnamefont
  {Wilczek}},\ }\href {\doibase 10.1103/PhysRevLett.53.722} {\bibfield
  {journal} {\bibinfo  {journal} {Phys. Rev. Lett.}\ }\textbf {\bibinfo
  {volume} {53}},\ \bibinfo {pages} {722} (\bibinfo {year} {1984})}\BibitemShut
  {NoStop}%
\bibitem [{\citenamefont {Bonderson}\ \emph {et~al.}(2011)\citenamefont
  {Bonderson}, \citenamefont {Gurarie},\ and\ \citenamefont
  {Nayak}}]{BonGurNay}%
  \BibitemOpen
  \bibfield  {author} {\bibinfo {author} {\bibfnamefont {P.}~\bibnamefont
  {Bonderson}}, \bibinfo {author} {\bibfnamefont {V.}~\bibnamefont {Gurarie}},
  \ and\ \bibinfo {author} {\bibfnamefont {C.}~\bibnamefont {Nayak}},\ }\href
  {\doibase 10.1103/PhysRevB.83.075303} {\bibfield  {journal} {\bibinfo
  {journal} {Phys. Rev. B}\ }\textbf {\bibinfo {volume} {83}},\ \bibinfo
  {pages} {075303} (\bibinfo {year} {2011})}\BibitemShut {NoStop}%
\bibitem [{\citenamefont {Wu}\ \emph {et~al.}(2014)\citenamefont {Wu},
  \citenamefont {Estienne}, \citenamefont {Regnault},\ and\ \citenamefont
  {Bernevig}}]{Braiding}%
  \BibitemOpen
  \bibfield  {author} {\bibinfo {author} {\bibfnamefont {Y.-L.}\ \bibnamefont
  {Wu}}, \bibinfo {author} {\bibfnamefont {B.}~\bibnamefont {Estienne}},
  \bibinfo {author} {\bibfnamefont {N.}~\bibnamefont {Regnault}}, \ and\
  \bibinfo {author} {\bibfnamefont {B.~A.}\ \bibnamefont {Bernevig}},\ }\href
  {\doibase 10.1103/PhysRevLett.113.116801} {\bibfield  {journal} {\bibinfo
  {journal} {Phys. Rev. Lett.}\ }\textbf {\bibinfo {volume} {113}},\ \bibinfo
  {pages} {116801} (\bibinfo {year} {2014})}\BibitemShut {NoStop}%
\bibitem [{\citenamefont {Manna}\ \emph {et~al.}(2019)\citenamefont {Manna},
  \citenamefont {Wildeboer},\ and\ \citenamefont {Nielsen}}]{Nielsen1}%
  \BibitemOpen
  \bibfield  {author} {\bibinfo {author} {\bibfnamefont {S.}~\bibnamefont
  {Manna}}, \bibinfo {author} {\bibfnamefont {J.}~\bibnamefont {Wildeboer}}, \
  and\ \bibinfo {author} {\bibfnamefont {A.~E.~B.}\ \bibnamefont {Nielsen}},\
  }\href {\doibase 10.1103/PhysRevB.99.045147} {\bibfield  {journal} {\bibinfo
  {journal} {Phys. Rev. B}\ }\textbf {\bibinfo {volume} {99}},\ \bibinfo
  {pages} {045147} (\bibinfo {year} {2019})}\BibitemShut {NoStop}%
\bibitem [{\citenamefont {Macaluso}\ \emph {et~al.}(2019)\citenamefont
  {Macaluso}, \citenamefont {Comparin}, \citenamefont {Mazza},\ and\
  \citenamefont {Carusotto}}]{macaluso}%
  \BibitemOpen
  \bibfield  {author} {\bibinfo {author} {\bibfnamefont {E.}~\bibnamefont
  {Macaluso}}, \bibinfo {author} {\bibfnamefont {T.}~\bibnamefont {Comparin}},
  \bibinfo {author} {\bibfnamefont {L.}~\bibnamefont {Mazza}}, \ and\ \bibinfo
  {author} {\bibfnamefont {I.}~\bibnamefont {Carusotto}},\ }\href {\doibase
  10.1103/PhysRevLett.123.266801} {\bibfield  {journal} {\bibinfo  {journal}
  {Phys. Rev. Lett.}\ }\textbf {\bibinfo {volume} {123}},\ \bibinfo {pages}
  {266801} (\bibinfo {year} {2019})}\BibitemShut {NoStop}%
\bibitem [{\citenamefont {Storni}\ and\ \citenamefont {Morf}(2011)}]{storni}%
  \BibitemOpen
  \bibfield  {author} {\bibinfo {author} {\bibfnamefont {M.}~\bibnamefont
  {Storni}}\ and\ \bibinfo {author} {\bibfnamefont {R.~H.}\ \bibnamefont
  {Morf}},\ }\href {\doibase 10.1103/PhysRevB.83.195306} {\bibfield  {journal}
  {\bibinfo  {journal} {Phys. Rev. B}\ }\textbf {\bibinfo {volume} {83}},\
  \bibinfo {pages} {195306} (\bibinfo {year} {2011})}\BibitemShut {NoStop}%
\bibitem [{\citenamefont {Moore}\ and\ \citenamefont {Read}(1991)}]{Read1}%
  \BibitemOpen
  \bibfield  {author} {\bibinfo {author} {\bibfnamefont {G.}~\bibnamefont
  {Moore}}\ and\ \bibinfo {author} {\bibfnamefont {N.}~\bibnamefont {Read}},\
  }\href {\doibase https://doi.org/10.1016/0550-3213(91)90407-O} {\bibfield
  {journal} {\bibinfo  {journal} {Nuclear Physics B}\ }\textbf {\bibinfo
  {volume} {360}},\ \bibinfo {pages} {362 } (\bibinfo {year}
  {1991})}\BibitemShut {NoStop}%
\bibitem [{\citenamefont {Glasser}\ \emph {et~al.}(2015)\citenamefont
  {Glasser}, \citenamefont {Cirac}, \citenamefont {Sierra},\ and\ \citenamefont
  {Nielsen}}]{Nielsen2}%
  \BibitemOpen
  \bibfield  {author} {\bibinfo {author} {\bibfnamefont {I.}~\bibnamefont
  {Glasser}}, \bibinfo {author} {\bibfnamefont {J.~I.}\ \bibnamefont {Cirac}},
  \bibinfo {author} {\bibfnamefont {G.}~\bibnamefont {Sierra}}, \ and\ \bibinfo
  {author} {\bibfnamefont {A.~E.~B.}\ \bibnamefont {Nielsen}},\ }\href
  {http://stacks.iop.org/1367-2630/17/i=8/a=082001} {\bibfield  {journal}
  {\bibinfo  {journal} {New Journal of Physics}\ }\textbf {\bibinfo {volume}
  {17}},\ \bibinfo {pages} {082001} (\bibinfo {year} {2015})}\BibitemShut
  {NoStop}%
\bibitem [{\citenamefont {Manna}\ \emph {et~al.}(2020)\citenamefont {Manna},
  \citenamefont {Pal}, \citenamefont {Wang},\ and\ \citenamefont
  {Nielsen}}]{Manna1}%
  \BibitemOpen
  \bibfield  {author} {\bibinfo {author} {\bibfnamefont {S.}~\bibnamefont
  {Manna}}, \bibinfo {author} {\bibfnamefont {B.}~\bibnamefont {Pal}}, \bibinfo
  {author} {\bibfnamefont {W.}~\bibnamefont {Wang}}, \ and\ \bibinfo {author}
  {\bibfnamefont {A.~E.~B.}\ \bibnamefont {Nielsen}},\ }\href {\doibase
  10.1103/PhysRevResearch.2.023401} {\bibfield  {journal} {\bibinfo  {journal}
  {Phys. Rev. Research}\ }\textbf {\bibinfo {volume} {2}},\ \bibinfo {pages}
  {023401} (\bibinfo {year} {2020})}\BibitemShut {NoStop}%
\bibitem [{\citenamefont {Hafezi}\ \emph {et~al.}(2007)\citenamefont {Hafezi},
  \citenamefont {S\o{}rensen}, \citenamefont {Demler},\ and\ \citenamefont
  {Lukin}}]{Lukin1}%
  \BibitemOpen
  \bibfield  {author} {\bibinfo {author} {\bibfnamefont {M.}~\bibnamefont
  {Hafezi}}, \bibinfo {author} {\bibfnamefont {A.~S.}\ \bibnamefont
  {S\o{}rensen}}, \bibinfo {author} {\bibfnamefont {E.}~\bibnamefont {Demler}},
  \ and\ \bibinfo {author} {\bibfnamefont {M.~D.}\ \bibnamefont {Lukin}},\
  }\href {\doibase 10.1103/PhysRevA.76.023613} {\bibfield  {journal} {\bibinfo
  {journal} {Phys. Rev. A}\ }\textbf {\bibinfo {volume} {76}},\ \bibinfo
  {pages} {023613} (\bibinfo {year} {2007})}\BibitemShut {NoStop}%
\bibitem [{\citenamefont {S\o{}rensen}\ \emph {et~al.}(2005)\citenamefont
  {S\o{}rensen}, \citenamefont {Demler},\ and\ \citenamefont {Lukin}}]{Lukin2}%
  \BibitemOpen
  \bibfield  {author} {\bibinfo {author} {\bibfnamefont {A.~S.}\ \bibnamefont
  {S\o{}rensen}}, \bibinfo {author} {\bibfnamefont {E.}~\bibnamefont {Demler}},
  \ and\ \bibinfo {author} {\bibfnamefont {M.~D.}\ \bibnamefont {Lukin}},\
  }\href {\doibase 10.1103/PhysRevLett.94.086803} {\bibfield  {journal}
  {\bibinfo  {journal} {Phys. Rev. Lett.}\ }\textbf {\bibinfo {volume} {94}},\
  \bibinfo {pages} {086803} (\bibinfo {year} {2005})}\BibitemShut {NoStop}%
\bibitem [{\citenamefont {Kapit}\ \emph {et~al.}(2012)\citenamefont {Kapit},
  \citenamefont {Ginsparg},\ and\ \citenamefont {Mueller}}]{kapitmueller}%
  \BibitemOpen
  \bibfield  {author} {\bibinfo {author} {\bibfnamefont {E.}~\bibnamefont
  {Kapit}}, \bibinfo {author} {\bibfnamefont {P.}~\bibnamefont {Ginsparg}}, \
  and\ \bibinfo {author} {\bibfnamefont {E.}~\bibnamefont {Mueller}},\ }\href
  {\doibase 10.1103/PhysRevLett.108.066802} {\bibfield  {journal} {\bibinfo
  {journal} {Phys. Rev. Lett.}\ }\textbf {\bibinfo {volume} {108}},\ \bibinfo
  {pages} {066802} (\bibinfo {year} {2012})}\BibitemShut {NoStop}%
\bibitem [{\citenamefont {Nielsen}\ \emph {et~al.}(2018)\citenamefont
  {Nielsen}, \citenamefont {Glasser},\ and\ \citenamefont
  {Rodr{\'\i}guez}}]{nielsen2018quasielectrons}%
  \BibitemOpen
  \bibfield  {author} {\bibinfo {author} {\bibfnamefont {A.~E.~B.}\
  \bibnamefont {Nielsen}}, \bibinfo {author} {\bibfnamefont {I.}~\bibnamefont
  {Glasser}}, \ and\ \bibinfo {author} {\bibfnamefont {I.~D.}\ \bibnamefont
  {Rodr{\'\i}guez}},\ }\href@noop {} {\bibfield  {journal} {\bibinfo  {journal}
  {New Journal of Physics}\ }\textbf {\bibinfo {volume} {20}},\ \bibinfo
  {pages} {033029} (\bibinfo {year} {2018})}\BibitemShut {NoStop}%
\bibitem [{\citenamefont {Raciunas}\ \emph {et~al.}(2018)\citenamefont
  {Raciunas}, \citenamefont {\"Unal}, \citenamefont {Anisimovas},\ and\
  \citenamefont {Eckardt}}]{eckardt}%
  \BibitemOpen
  \bibfield  {author} {\bibinfo {author} {\bibfnamefont {M.}~\bibnamefont
  {Raciunas}}, \bibinfo {author} {\bibfnamefont {F.~N.}\ \bibnamefont
  {\"Unal}}, \bibinfo {author} {\bibfnamefont {E.}~\bibnamefont {Anisimovas}},
  \ and\ \bibinfo {author} {\bibfnamefont {A.}~\bibnamefont {Eckardt}},\ }\href
  {\doibase 10.1103/PhysRevA.98.063621} {\bibfield  {journal} {\bibinfo
  {journal} {Phys. Rev. A}\ }\textbf {\bibinfo {volume} {98}},\ \bibinfo
  {pages} {063621} (\bibinfo {year} {2018})}\BibitemShut {NoStop}%
\bibitem [{\citenamefont {Kitaev}(2006)}]{Kitaev1}%
  \BibitemOpen
  \bibfield  {author} {\bibinfo {author} {\bibfnamefont {A.}~\bibnamefont
  {Kitaev}},\ }\href {\doibase https://doi.org/10.1016/j.aop.2005.10.005}
  {\bibfield  {journal} {\bibinfo  {journal} {Annals of Physics}\ }\textbf
  {\bibinfo {volume} {321}},\ \bibinfo {pages} {2} (\bibinfo {year}
  {2006})}\BibitemShut {NoStop}%
\bibitem [{\citenamefont {Kitaev}(2003)}]{Kitaev2}%
  \BibitemOpen
  \bibfield  {author} {\bibinfo {author} {\bibfnamefont {A.}~\bibnamefont
  {Kitaev}},\ }\href {\doibase https://doi.org/10.1016/S0003-4916(02)00018-0}
  {\bibfield  {journal} {\bibinfo  {journal} {Annals of Physics}\ }\textbf
  {\bibinfo {volume} {303}},\ \bibinfo {pages} {2} (\bibinfo {year}
  {2003})}\BibitemShut {NoStop}%
\bibitem [{\citenamefont {Vidal}\ \emph
  {et~al.}(2009{\natexlab{a}})\citenamefont {Vidal}, \citenamefont {Dusuel},\
  and\ \citenamefont {Schmidt}}]{Vidal4}%
  \BibitemOpen
  \bibfield  {author} {\bibinfo {author} {\bibfnamefont {J.}~\bibnamefont
  {Vidal}}, \bibinfo {author} {\bibfnamefont {S.}~\bibnamefont {Dusuel}}, \
  and\ \bibinfo {author} {\bibfnamefont {K.~P.}\ \bibnamefont {Schmidt}},\
  }\href {\doibase 10.1103/PhysRevB.79.033109} {\bibfield  {journal} {\bibinfo
  {journal} {Phys. Rev. B}\ }\textbf {\bibinfo {volume} {79}},\ \bibinfo
  {pages} {033109} (\bibinfo {year} {2009}{\natexlab{a}})}\BibitemShut
  {NoStop}%
\bibitem [{\citenamefont {Vidal}\ \emph
  {et~al.}(2009{\natexlab{b}})\citenamefont {Vidal}, \citenamefont {Thomale},
  \citenamefont {Schmidt},\ and\ \citenamefont {Dusuel}}]{Vidal2}%
  \BibitemOpen
  \bibfield  {author} {\bibinfo {author} {\bibfnamefont {J.}~\bibnamefont
  {Vidal}}, \bibinfo {author} {\bibfnamefont {R.}~\bibnamefont {Thomale}},
  \bibinfo {author} {\bibfnamefont {K.~P.}\ \bibnamefont {Schmidt}}, \ and\
  \bibinfo {author} {\bibfnamefont {S.}~\bibnamefont {Dusuel}},\ }\href
  {\doibase 10.1103/PhysRevB.80.081104} {\bibfield  {journal} {\bibinfo
  {journal} {Phys. Rev. B}\ }\textbf {\bibinfo {volume} {80}},\ \bibinfo
  {pages} {081104(R)} (\bibinfo {year} {2009}{\natexlab{b}})}\BibitemShut
  {NoStop}%
\bibitem [{\citenamefont {Dusuel}\ \emph {et~al.}(2011)\citenamefont {Dusuel},
  \citenamefont {Kamfor}, \citenamefont {Or\'us}, \citenamefont {Schmidt},\
  and\ \citenamefont {Vidal}}]{Vidal3}%
  \BibitemOpen
  \bibfield  {author} {\bibinfo {author} {\bibfnamefont {S.}~\bibnamefont
  {Dusuel}}, \bibinfo {author} {\bibfnamefont {M.}~\bibnamefont {Kamfor}},
  \bibinfo {author} {\bibfnamefont {R.}~\bibnamefont {Or\'us}}, \bibinfo
  {author} {\bibfnamefont {K.~P.}\ \bibnamefont {Schmidt}}, \ and\ \bibinfo
  {author} {\bibfnamefont {J.}~\bibnamefont {Vidal}},\ }\href {\doibase
  10.1103/PhysRevLett.106.107203} {\bibfield  {journal} {\bibinfo  {journal}
  {Phys. Rev. Lett.}\ }\textbf {\bibinfo {volume} {106}},\ \bibinfo {pages}
  {107203} (\bibinfo {year} {2011})}\BibitemShut {NoStop}%
\bibitem [{\citenamefont {Zarei}(2019)}]{zarei}%
  \BibitemOpen
  \bibfield  {author} {\bibinfo {author} {\bibfnamefont {M.~H.}\ \bibnamefont
  {Zarei}},\ }\href {\doibase 10.1103/PhysRevB.100.125159} {\bibfield
  {journal} {\bibinfo  {journal} {Phys. Rev. B}\ }\textbf {\bibinfo {volume}
  {100}},\ \bibinfo {pages} {125159} (\bibinfo {year} {2019})}\BibitemShut
  {NoStop}%
\bibitem [{\citenamefont {Greplova}\ \emph {et~al.}(2020)\citenamefont
  {Greplova}, \citenamefont {Valenti}, \citenamefont {Boschung}, \citenamefont
  {Sch{\"a}fer}, \citenamefont {L{\"o}rch},\ and\ \citenamefont
  {Huber}}]{greplova}%
  \BibitemOpen
  \bibfield  {author} {\bibinfo {author} {\bibfnamefont {E.}~\bibnamefont
  {Greplova}}, \bibinfo {author} {\bibfnamefont {A.}~\bibnamefont {Valenti}},
  \bibinfo {author} {\bibfnamefont {G.}~\bibnamefont {Boschung}}, \bibinfo
  {author} {\bibfnamefont {F.}~\bibnamefont {Sch{\"a}fer}}, \bibinfo {author}
  {\bibfnamefont {N.}~\bibnamefont {L{\"o}rch}}, \ and\ \bibinfo {author}
  {\bibfnamefont {S.~D.}\ \bibnamefont {Huber}},\ }\href {\doibase
  10.1088/1367-2630/ab7771} {\bibfield  {journal} {\bibinfo  {journal} {New
  Journal of Physics}\ }\textbf {\bibinfo {volume} {22}},\ \bibinfo {pages}
  {045003} (\bibinfo {year} {2020})}\BibitemShut {NoStop}%
\bibitem [{\citenamefont {Weitenberg}\ \emph {et~al.}(2011)\citenamefont
  {Weitenberg}, \citenamefont {Endres}, \citenamefont {Sherson}, \citenamefont
  {Cheneau}, \citenamefont {Schau{\ss}}, \citenamefont {Fukuhara},
  \citenamefont {Bloch},\ and\ \citenamefont {Kuhr}}]{weitenberg}%
  \BibitemOpen
  \bibfield  {author} {\bibinfo {author} {\bibfnamefont {C.}~\bibnamefont
  {Weitenberg}}, \bibinfo {author} {\bibfnamefont {M.}~\bibnamefont {Endres}},
  \bibinfo {author} {\bibfnamefont {J.~F.}\ \bibnamefont {Sherson}}, \bibinfo
  {author} {\bibfnamefont {M.}~\bibnamefont {Cheneau}}, \bibinfo {author}
  {\bibfnamefont {P.}~\bibnamefont {Schau{\ss}}}, \bibinfo {author}
  {\bibfnamefont {T.}~\bibnamefont {Fukuhara}}, \bibinfo {author}
  {\bibfnamefont {I.}~\bibnamefont {Bloch}}, \ and\ \bibinfo {author}
  {\bibfnamefont {S.}~\bibnamefont {Kuhr}},\ }\href@noop {} {\bibfield
  {journal} {\bibinfo  {journal} {Nature}\ }\textbf {\bibinfo {volume} {471}},\
  \bibinfo {pages} {319} (\bibinfo {year} {2011})}\BibitemShut {NoStop}%
\bibitem [{\citenamefont {Chen}\ \emph {et~al.}(2019)\citenamefont {Chen},
  \citenamefont {Zhang}, \citenamefont {Zhang}, \citenamefont {Liu},
  \citenamefont {Kou}, \citenamefont {Sun},\ and\ \citenamefont
  {Zhang}}]{Toric1}%
  \BibitemOpen
  \bibfield  {author} {\bibinfo {author} {\bibfnamefont {T.}~\bibnamefont
  {Chen}}, \bibinfo {author} {\bibfnamefont {S.}~\bibnamefont {Zhang}},
  \bibinfo {author} {\bibfnamefont {Y.}~\bibnamefont {Zhang}}, \bibinfo
  {author} {\bibfnamefont {Y.}~\bibnamefont {Liu}}, \bibinfo {author}
  {\bibfnamefont {S.-P.}\ \bibnamefont {Kou}}, \bibinfo {author} {\bibfnamefont
  {H.}~\bibnamefont {Sun}}, \ and\ \bibinfo {author} {\bibfnamefont
  {X.}~\bibnamefont {Zhang}},\ }\href@noop {} {\bibfield  {journal} {\bibinfo
  {journal} {Nature communications}\ }\textbf {\bibinfo {volume} {10}},\
  \bibinfo {pages} {1557} (\bibinfo {year} {2019})}\BibitemShut {NoStop}%
\bibitem [{\citenamefont {Luo}\ \emph {et~al.}(2018)\citenamefont {Luo},
  \citenamefont {Li}, \citenamefont {Li}, \citenamefont {Hung}, \citenamefont
  {Wan}, \citenamefont {Peng},\ and\ \citenamefont {Du}}]{Toric2}%
  \BibitemOpen
  \bibfield  {author} {\bibinfo {author} {\bibfnamefont {Z.}~\bibnamefont
  {Luo}}, \bibinfo {author} {\bibfnamefont {J.}~\bibnamefont {Li}}, \bibinfo
  {author} {\bibfnamefont {Z.}~\bibnamefont {Li}}, \bibinfo {author}
  {\bibfnamefont {L.-Y.}\ \bibnamefont {Hung}}, \bibinfo {author}
  {\bibfnamefont {Y.}~\bibnamefont {Wan}}, \bibinfo {author} {\bibfnamefont
  {X.}~\bibnamefont {Peng}}, \ and\ \bibinfo {author} {\bibfnamefont
  {J.}~\bibnamefont {Du}},\ }\href@noop {} {\bibfield  {journal} {\bibinfo
  {journal} {Nature physics}\ }\textbf {\bibinfo {volume} {14}},\ \bibinfo
  {pages} {160} (\bibinfo {year} {2018})}\BibitemShut {NoStop}%
\bibitem [{\citenamefont {Pachos}\ \emph {et~al.}(2009)\citenamefont {Pachos},
  \citenamefont {Wieczorek}, \citenamefont {Schmid}, \citenamefont {Kiesel},
  \citenamefont {Pohlner},\ and\ \citenamefont {Weinfurter}}]{Toric3}%
  \BibitemOpen
  \bibfield  {author} {\bibinfo {author} {\bibfnamefont {J.~K.}\ \bibnamefont
  {Pachos}}, \bibinfo {author} {\bibfnamefont {W.}~\bibnamefont {Wieczorek}},
  \bibinfo {author} {\bibfnamefont {C.}~\bibnamefont {Schmid}}, \bibinfo
  {author} {\bibfnamefont {N.}~\bibnamefont {Kiesel}}, \bibinfo {author}
  {\bibfnamefont {R.}~\bibnamefont {Pohlner}}, \ and\ \bibinfo {author}
  {\bibfnamefont {H.}~\bibnamefont {Weinfurter}},\ }\href@noop {} {\bibfield
  {journal} {\bibinfo  {journal} {New Journal of Physics}\ }\textbf {\bibinfo
  {volume} {11}},\ \bibinfo {pages} {083010} (\bibinfo {year}
  {2009})}\BibitemShut {NoStop}%
\end{thebibliography}%

\end{document}